\documentclass[twoside,12pt]{article}
\usepackage{epsfig}

\def\Journal#1#2#3#4{{#1} {#2} (#4) #3 }

\def\NPA{{\em Nucl. Phys.} A}

\def\PLB{{\em Phys. Lett.} B}

\newcommand{\be}{\begin{equation}}
\newcommand{\ee}{\end{equation}}
\newcommand{\bea}{\begin{eqnarray}}
\newcommand{\eea}{\end{eqnarray}}

\begin{document}

\title{ \vspace{1cm} Electromagnetic Transition Form Factors of Mesons}
\author{C.\ Terschl\"usen$^{1}$, S.\ Leupold$^1$\\
\\
$^1$Department of Physics and Astronomy, Uppsala University, Sweden}
\maketitle
\begin{abstract} 
Using a counting scheme which treats pseudoscalar and vector mesons on equal footing, the decays of the narrow light vector mesons $\omega$ and $\phi$ into a dilepton and a pseudoscalar $\pi^0$-meson or $\eta$-meson, respectively, are calculated. Thereby, all required parameters could be determined by other reactions so that one has predictive power for the considered decays. The calculated partial decay widths are in very good agreement with the experimental data.
\end{abstract}

\section{Introduction}
\begin{figure}
 \centering
 \epsfig{file=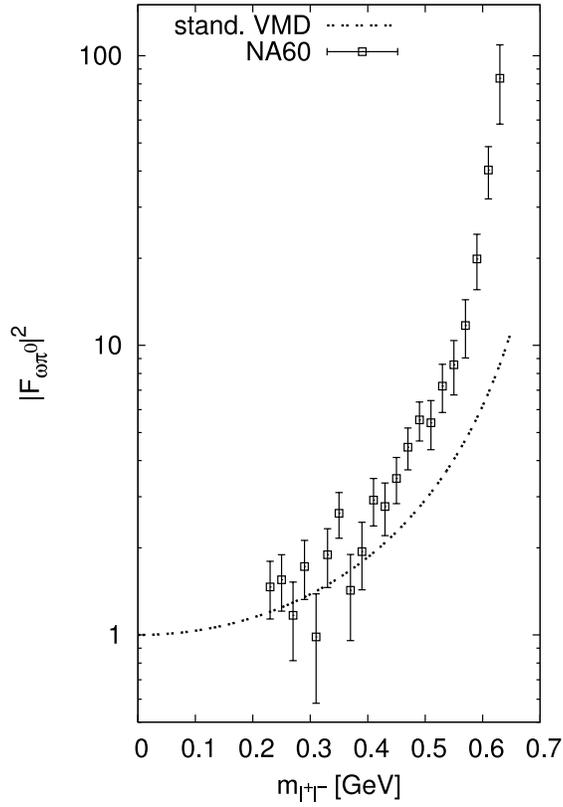, scale=0.45}
 \caption{$\omega \rightarrow \pi^0$ transition form factor calculated with VMD (dot-dashed) in comparison to form-factor data taken by the NA60 collaboration for the decay into a dimuon \cite{NA60}.}
 \label{fig:omegapinurVMD}
\end{figure}

Probing hadrons by electromagnetic reactions strongly involves vector mesons according to the vector-meson dominance conjecture \cite{Sakurai}. Although the standard vector-meson-dominance model (VMD) is quite successful to describe these kinds of reactions, it is not able to describe all of them. In the following, we will study transition form factors, i.e. decays of a vector meson into a pseudoscalar meson and a dilepton \cite{vectorpaper}. In fig. \ref{fig:omegapinurVMD}, the $\omega \rightarrow \pi^0$ transition form factor calculated with VMD is plotted in comparison to form-factor data taken by the NA60 collaboration for the decay into a dimuon \cite{NA60}. Obviously, VMD fails to describe the data. \\
Due to the running coupling constant in QCD, perturbative QCD can only be used for high energies. For low energies, this is not possible. A possible solution are effective theories which use hadrons as relevant degrees of freedom. In the lowest energy region, the effective theory called Chiral perturbation theory (ChPT) \cite{GL,Scherer} is such an effective theory. The only active degrees of freedom are the pseudoscalar Goldstone bosons ($\pi$, $K$, $\eta$). All other mesons, in particular the vector mesons ($\rho$, $\omega$, $K^\ast$, $\phi$) are treated as heavy. Therefore, ChPT is not applicable for the energy range of the hadronic resonances. So far, there have only been phenomenologically successful models as, e.g., VMD for this energy range. Our final aim is to develop an effective theory for this range. \\
We use a new counting scheme for both the Goldstone bosons and the light vector mesons proposed in \cite{CS, dreivector}. According to this scheme, the masses of both the Goldstone bosons and the light vector mesons are treated as soft, i.e. of the order of a typical momentum $q$. If one describes decays, all involved momenta will be smaller than the mass of the decaying meson and, thus, also of the order of $q$,
\begin{eqnarray}
 m_V, \, m_P, \, \partial_\mu \sim q.
\end{eqnarray}
The restriction to these mesons can be justified by the hadrogenesis conjecture, i.e. all other low-lying mesons are considered as dynamically generated from interactions of Goldstone bosons and light vector mesons\footnote{The systematic inclusion of the $\eta$ singlet is part of our ongoing work.}.

In ChPT, the range of applicability, i.e. the range for $q$ is limited (on tree level) by the not-considered mesons, in practice by $m_V$, and (for loops) by the scale $4 \pi f$, where $f$ denotes the pion decay constant. In the scheme of \cite{CS}, where vector mesons are included and where two-particle reducible diagrams (rescattering processes) are resummed, it is suggestive that the range of applicability can be pushed to larger energies. Next-to-leading-order calculations are necessary to assess this proposition. At present, phenomenological consequences are worked out in leading order and compared to data.

\section{Leading-order Lagrangian}
Using the counting scheme \cite{CS}, one can determine the leading-order chiral Lagrangian for the decay $V \rightarrow P \gamma^\ast$,
\begin{eqnarray}
 \mathcal{L}_{\rm indir.} = &&- \, \frac{1}{16 f} \, h_A \, \varepsilon^{\mu\nu\alpha\beta} \, {\rm tr}\left\{ \left[ V_{\mu\nu}, \partial^\tau V_{\tau\alpha} \right]_+ \partial_\beta \Phi \right\} - \, \frac{1}{16 f} \, b_A \, \varepsilon^{\mu\nu\alpha\beta} \, {\rm tr}\left\{ \left[ V_{\mu\nu}, V_{\alpha, \beta} \right]_+ \left[\Phi, \chi_0 \right]_+ \right\} \nonumber \\
 && - \, \frac{e_V m_V}{4} \, \left\{ V^{\mu\nu} Q \right\} \partial_\mu A_\nu . \label{LOL}
\end{eqnarray}
Thereby, the matrix $V_{\mu\nu}$ describes the vector mesons represented by antisymmetric tensor fields,
\begin{eqnarray}
 V_{\mu\nu} = \left(\begin{array}{ccc}
                \rho^0_{\mu\nu} + \omega_{\mu\nu} & \sqrt{2} \rho^+_{\mu\nu} & \sqrt{2} K^+_{\mu\nu} \\
                \sqrt{2} \rho^-_{\mu\nu} & - \rho^0_{\mu\nu} + \omega_{\mu\nu} & \sqrt{2} K^0_{\mu\nu} \\
		\sqrt{2} K^-_{\mu\nu} & \sqrt{2} \bar{K}^0_{\mu\nu} & \sqrt{2} \phi_{\mu\nu}
               \end{array} \right) \,,
\end{eqnarray}
and $\Phi$ the Goldstone bosons,
\begin{eqnarray}
 \Phi = \left(\begin{array}{ccc}
           \pi^0 + \frac{1}{\sqrt{3}} \eta & \sqrt{2} \pi^+ & \sqrt{2} K^+ \\
	   \sqrt{2} \pi^- & - \pi^0 + \frac{1}{\sqrt{3}} \eta & \sqrt{2} K^0 \\
	   \sqrt{2} K^- & \sqrt{2} \bar{K}^0 & - \frac{2}{\sqrt{3}} \eta 
          \end{array}\right)\,.
\end{eqnarray} 
Furthermore, we have introduced the mass matrix $\chi_0 = {\rm diag}\left( m_\pi^2, m_\pi^2, 2 m_K^2 - m_\pi^2 \right)$ and the quark-charge matrix $Q = {\rm diag}\left( 2/3, -1/3, -1/3 \right)$. $A_\mu$ describes the photon field.  \\
The leading-order Lagrangian (\ref{LOL}) allows only indirect decays, i.e. decays via a virtual vector meson. The first two terms proportional to the parameters $h_A$ and $b_A$ describe the decay of a vector meson into a virtual vector meson and a Goldstone boson and the third term describes the decay of the virtual vector meson into a (real or virtual) photon. The decay of the photon into a dilepton is described by usual QED.\\
To determine the uncertainties of our method, we use as a very rough estimate one particular next-to-leading-order term of the Lagrangian,
\begin{eqnarray}
 \mathcal{L}_{\rm dir.} = - \, \frac{1}{4 f m_V} \, e_A \, \varepsilon^{\mu\nu\alpha\beta} \, {\rm tr} \left\{ \left[ Q, \partial^\tau V_{\tau \alpha} \right]_+ \partial_\beta \Phi \right\} \partial_\mu A_\nu \label{NLOL}.
\end{eqnarray}
This terms describes the direct decay of a vector meson into a Goldstone boson and a (real or virtual) photon. \\
The open parameters $h_A$, $b_A$ and $e_A$ can be fixed by fitting the partial decay widths for the two-body decays $V \rightarrow P \gamma$ to the available experimental data. We fixed two parameter sets, parameter set (P1) with $e_A = 0$ and $(h_A, b_A) = (2.32, 0.27)$ which describes the leading-order calculation and parameter set (P2) with $e_A = 0.015$ and $(h_A,b_A) = (2.10, 0.19)$ which includes the particular next-to-leading-order term.\\
For the decays into dileptons no additional parameters are needed so that we have predictive power for these decays. Hence, there is no additional fitting to experimental data included in the results which will be presented in the next section.

\section{Results}
\subsection{Decay $\omega \rightarrow \pi^0 l^+ l^-$}
\begin{figure}[th]
\centering
  \begin{minipage}{.45\textwidth}
    \begin{center}  
      \epsfig{file=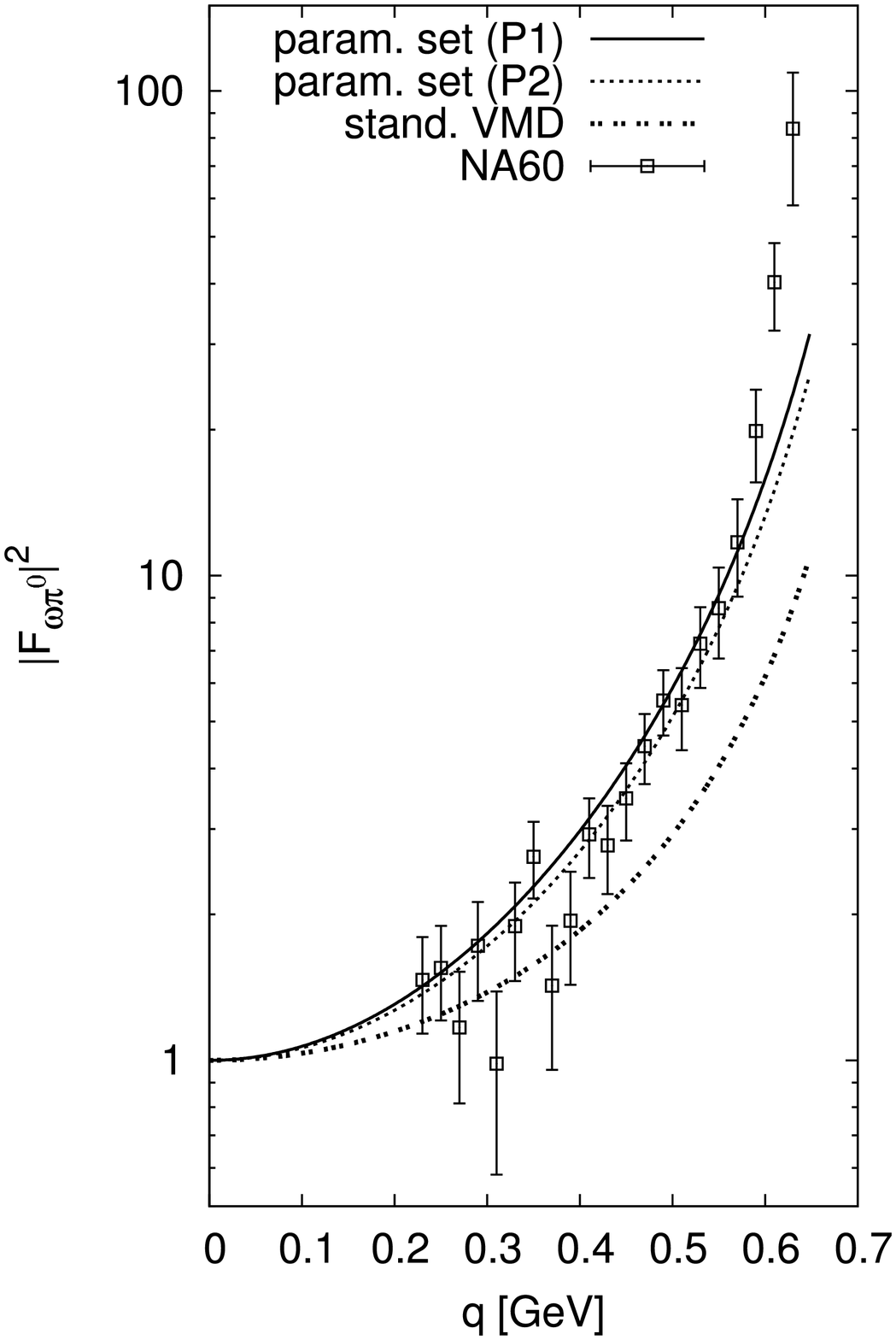, scale=0.45}
    \end{center}
  \end{minipage}
  \hfill
  \begin{minipage}{.45\textwidth}
    \begin{center}  
      \epsfig{file=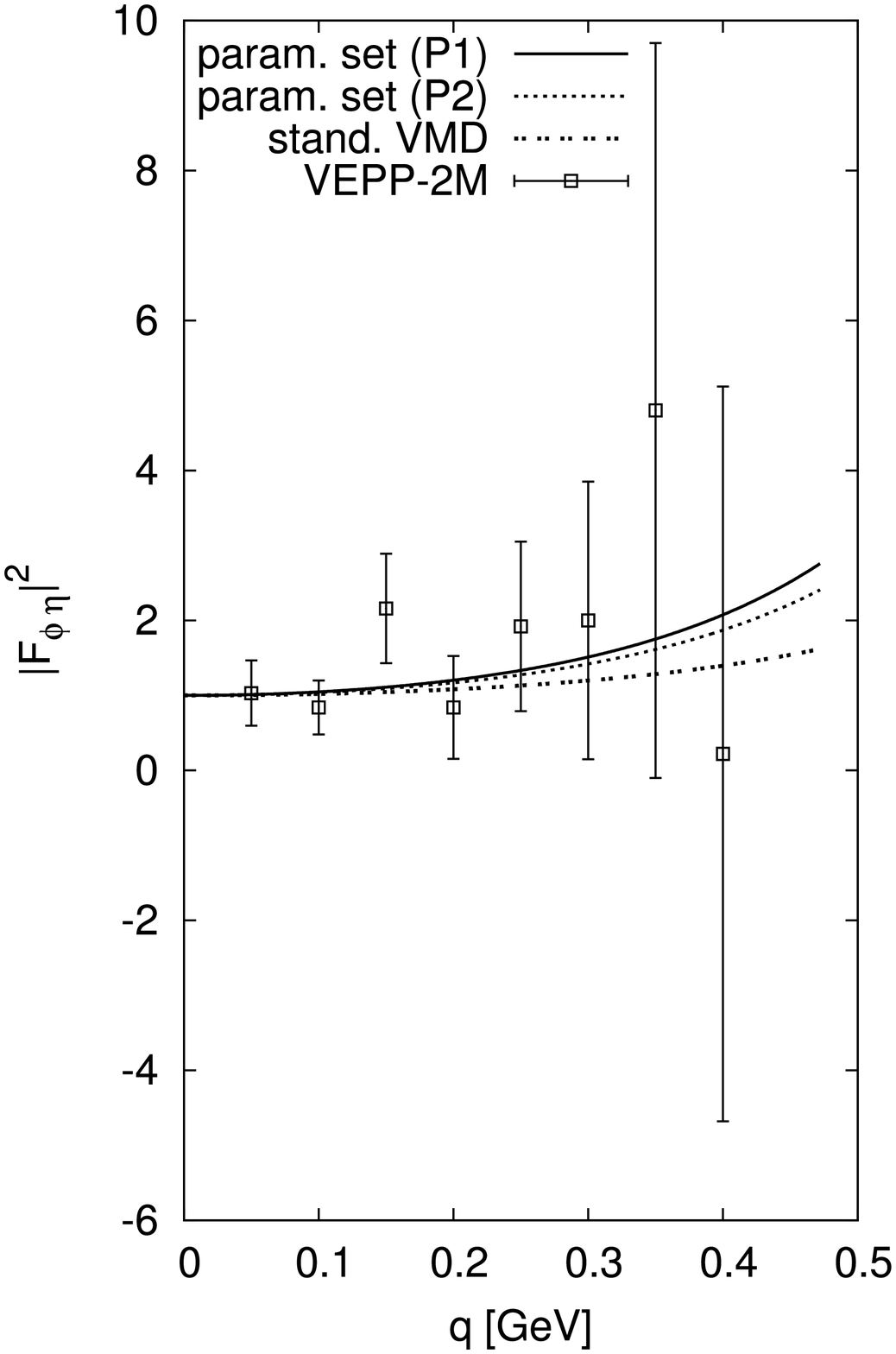, scale=0.45}
    \end{center}
  \end{minipage}
  \hfill
 \caption{\textbf{Left:} Form factor for the decay $\omega \rightarrow \pi^0 l^+ l^-$ compared to the dimuon data taken by the NA60 collaboration \cite{NA60}. \textbf{Right:} Form factor for the decay $\phi \rightarrow \eta l^+ l^-$ compared to the dielectron data taken with the VEPP-2M collider \cite{VEPP}. The solid lines describe the form factors calculated with parameter set (P1), the dotted lines with parameter set (P2) and the dot-dashed lines describe the VMD calculations.}
 \label{fig:omegapi}
\end{figure}

Due to isospin conservation, the decay $\omega \rightarrow \pi^0 l^+ l^-$ is only possible via a virtual $\rho^0$ meson. Thus, the transition form factor calculated with VMD equals
\begin{eqnarray}
 F_{\omega \pi^0}^{\rm VMD}(q) = \frac{m_\rho^2}{m_\rho^2-q^2} \label{FfVMDomegapi}
\end{eqnarray}
whereby $q^2$ is the square of the invariant mass of the dilepton. The form factor calculated with the Lagrangians (\ref{LOL}, \ref{NLOL}) contains both a term of VMD type and a constant term,
\begin{eqnarray}
 F_{\omega \pi^0}(q) = g_{\omega \pi^0} \, \frac{m_\rho^2}{m_\rho^2-q^2} \, + \left(1-g_{\omega \pi^0}\right)
\end{eqnarray}
with the abbreviation
\begin{eqnarray}
 g_{\omega \pi^0} = \frac{h_A \left( m_\rho^2 + m_\omega^2 \right) - 8 b_A m_\pi^2}{\left( h_A + 4 \, \frac{e_A}{e_V} \right) m_\omega^2 - 8 b_A m_\pi^2} = 2.01 \pm 0.24.
\end{eqnarray}
Note that for $g_{\omega \pi^0} = 2$ one obtains a particularly simple form 
\begin{eqnarray}
 F_{\omega \pi^0}(q) = \frac{m_\rho^2 + q^2}{m_\rho^2 - q^2} 
\end{eqnarray}
which is very different from (\ref{FfVMDomegapi}). \\
On the left-hand side of Fig. \ref{fig:omegapi}, this form factor is plotted \cite{vectorpaper} in comparison to the NA60 data \cite{NA60} and the VMD form factor. As already mentioned in the introduction, the VMD form factor fails to describe the data. Our calculations miss only the last three data points. Additionally, the partial decay widths for the decays into dimuon and dielectron agree very well with the experimental data \cite{vectorpaper, PDG},
\begin{eqnarray}
 \Gamma_{\omega \rightarrow \pi^0 \mu^+ \mu^-} &=& (9.85 \pm 0.58) \cdot 10^{-7} \, {\rm GeV}, \\
 \Gamma_{\omega \rightarrow \pi^0 \mu^+ \mu^-}^{\rm exp.} &=& (8.15 \pm 2.13) \cdot 10^{-7} \, {\rm GeV}, \\
 \Gamma_{\omega \rightarrow \pi^0 e^+ e^-} &=& (6.93 \pm 0.09) \cdot 10^{-6} \, {\rm GeV}, \\
 \Gamma_{\omega \rightarrow \pi^0 e^+ e^-}^{\rm exp.} &=& (6.54 \pm 0.54) \cdot 10^{-6} \, {\rm GeV}.
\end{eqnarray}
A differential measurement of $\omega \rightarrow \pi^0 \, e^+ e^-$ is planned at WASA-at-COSY.

\subsection{Decay $\phi \rightarrow \eta l^+ l^-$}
In agreement with isospin conservation and suppression of a decay via a virtual $\omega$ meson due to the OZI rule, the decay $\phi \rightarrow \eta l^+ l^-$ happens via a virtual $\phi$ meson in leading order. Our form factor for the $\phi \rightarrow \eta$ transition includes again an additional constant term,
\begin{eqnarray}
 F_{\phi \eta}(q) &=& g_{\phi \eta} \, \frac{m_\phi^2}{m_\phi^2 - q^2} \, + \left( 1- g_{\phi\eta} \right) \\
 {\rm with } \hspace{5mm} g_{\phi \eta} &=& \frac{2 h_A m_\phi^2 - 8 b_A \left(2 m_K^2 - m_\pi^2 \right)}{\left( 4 \, \frac{e_A}{e_V} \, \frac{m_\phi^2}{m_V^2} \, + h_A \right) m_\phi^2 - 8 b_A \left( 2 m_K^2 - m_\pi^2 \right)} = 2.74 \pm 0.87 \,,
\end{eqnarray}
in comparison to the standard VMD form factor, 
\begin{eqnarray}
  F_{\phi \eta}^{\rm VMD}(q) &=& \frac{m_\phi^2}{m_\phi^2 - q^2} \,.
\end{eqnarray}
Both form factors are plotted on the right-hand side of Fig. \ref{fig:omegapi} \cite{vectorpaper} in comparison to form-factor data taken at the VEPP-2M collider for the decay into a dielectron \cite{VEPP}. Unfortunately, the error bars are relatively large so that no assessment is possible whether our calculation or the VMD form factor describes the data better. However, a deviation between both calculations is visible. In the near future, data with improved quality are expected from the KLOE collaboration.\\
The partial decay width for the decay into a dielectron is in agreement with the experimental data \cite{vectorpaper, PDG},
\begin{eqnarray}
 \Gamma_{\phi \rightarrow \eta e^+ e^-} &=& (4.64 \pm 0.26) \cdot 10^{-7} \, {\rm GeV}, \\
 \Gamma_{\phi \rightarrow \eta e^+ e^-}^{\rm exp.} &=& (4.90 \pm 0.47) \cdot 10^{-7} \, {\rm GeV}.
\end{eqnarray}
There is no experimental data available for the decay into a dimuon so that our calculation has to be seen as a prediction \cite{vectorpaper},
\begin{eqnarray}
 \Gamma_{\phi \rightarrow \eta \mu^+ \mu^-} = (2.75 \pm 0.29) \cdot 10^{-8} \, {\rm GeV}.
\end{eqnarray}

\section{Summary and outlook}
Using the counting scheme proposed in \cite{CS}, the calculated form factors for the transitions $\omega \rightarrow \pi^0$ and $\phi \rightarrow \eta$ describe the available form-factor data well. For the transition $\omega \rightarrow \pi^0$, the data is much better described than with VMD. Additionally, the calculated partial decay widths for the decays into dileptons agree very well with the available experimental data. \\
To improve our method, a systematic inclusion of the $\eta'$ meson as an additional degree of freedom and next-to-leading-order calculations are necessary.

\end{document}